\documentclass[twoside]{elsart}
\usepackage{amssymb,amsbsy,amsmath}
\usepackage{times}

\newtheorem{lemma}{Lemma}
\newtheorem{Def}{Definition}
\newtheorem{theorem}{Theorem}
\newcommand{\eqdef}{\stackrel{\rm def}{=}}
\newcommand{\EinsOp}
           {\;\smash{\raisebox{0.0ex}{$\!\!\!\mbox{1}
            \hspace{-0.4ex}\rule[0.0ex]{0.06ex}{1.60ex}$}}}
\begin{document}
%---------- Journal-----------------------------------------
%
\journal{Discrete Applied Mathematics}
\begin{frontmatter}
\title{On cyclically shifted strings}
 
\author{K. B\"arwinkel, H.-J. Schmidt\thanksref{HJS}}
\author{$\!\!$, J. Schnack}
\address{Universit\"at Osnabr\"uck, Fachbereich Physik \\ 
         Barbarastr. 7, 49069 Osnabr\"uck, Germany}
\thanks[HJS]{corresponding author: hschmidt\char'100uos.de,\\ http://www.physik.uni-osnabrueck.de/makrosysteme/}

\begin{abstract}
\noindent
If a string is cyclically shifted it will re--appear after a certain number
of shifts, which will be called its {\it order}. 
We solve the problem of how many strings exist with a given order.
This problem arises in the context of quantum mechanics of spin systems.
\end{abstract}
\end{frontmatter}
\raggedbottom
%%%%%%%%%%%%%%%%%%%%%%%%%%%%%%%%%%%%%%%%%%%%%%%%%%%%%%%%%%%%%%%%%%%%%%%%
\section{Introduction and definitions}
Let ${\mathcal S}(A,N)$ denote the set of strings 
$a=\langle a_1,\ldots,a_N \rangle$ of natural numbers 
$a_n\in\left\{0,\ldots ,A-1\right\}$. There are exactly $A^N$ such strings.
For any $a\in {\mathcal S}(A,N)$ let $\Sigma(a)\eqdef \sum_{n=0}^N a_n$ and
$T(a)\eqdef \langle a_N,a_1,a_2,\ldots,a_{N-1}\rangle$.
$T$ is the cyclic shift operator. If $T^n$  denotes the $n$th power of $T$, $n\in{\Bbb N}$,
it follows that $T^N=T^0=\EinsOp_{{\mathcal S}(A,N)}$.\\
We consider two equivalence relations on ${\mathcal S}(A,N)$. For $a,b\in{\mathcal S}(A,N)$ we define
\begin{equation}
a\sim b \Leftrightarrow \Sigma(a)=\Sigma(b)
\end{equation}
and
\begin{equation}
a\approx  b \Leftrightarrow a=T^n(b) \mbox{ for some } n\in{\Bbb N}.
\end{equation}
Obviously, $a\approx b$ implies $a \sim b$ since the sum of the numbers in a string is 
invariant under permutations.\\
The aim of this article is to analyze the structure of the equivalence classes of strings with %%@
respect to $\sim$ and $\approx$. The main question will be: How many $\approx$-equivalence %%@
classes of a given size exist? Or: How many $\approx$-equivalence classes of a given size %%@
exist which are contained in a certain $\sim$-equivalence class? This problem can, of %%@
course, be solved in a straight-forward manner for any given $A$ and $N$, either by hand or by %%@
means of a simple computer program. We are rather seeking explicit formulae which answer the %%@
above questions.\\
The problem arises in the context of quantum mechanics of spin rings with a cyclically %%@
symmetric coupling between the $N$ individual spins. Any individual spin can assume $A$ %%@
different states and the total system can assume $A^N$ different states. More precisely:
The total Hilbert space of the problem possesses an orthonormal basis of product states parametrized by the set
${\mathcal S}(A,N)$. According to the symmetries of the problem it is possible to split the total %%@
Hilbert space into a sum of orthogonal subspaces which are invariant under the Hamiltonian of the problem. 
These subspaces are closely connected to the equivalence classes of strings %%@
defined above. For more details %%@
see \cite{Kou97,Kou98,BSS99}.

%-----------------------------------------------------------
\section{Strings with constant sum}

For any $a\in{\mathcal S}(A,N)$ we denote the equivalence class of strings having the same sum by
\begin{equation}
[a]_\sim \eqdef {\mathcal S}(A,N,M) \quad\mbox{ where } M\eqdef \Sigma(a).
\end{equation}
Obviously, ${\mathcal S}(A,N)$ is a disjoint union
\begin{equation}
{\mathcal S}(A,N) =\bigcup_{M=0\ldots N(A-1)}{\mathcal S}(A,N,M) 
\end{equation}
and the total number of strings satisfies
\begin{equation}
\left|{\mathcal S}(A,N)\right|=A^N =\sum_{M=0\ldots N(A-1)}\left|{\mathcal S}(A,N,M)\right|. 
\end{equation}
The problem of determining the number of strings with a constant sum 
$\left|{\mathcal S}(A,N,M)\right|$ is equivalent to the problem of calculating the probability %%@
distribution of the sum of $N$ independent, finite, uniformly distributed random variables. An %%@
example would be the probability of scoring the sum $M$ in a throw with $N$ dice with $A$ %%@
faces. Geometrically, this is the problem of how many lattice points are met if you cut a %%@
hypercube containing $A^N$ lattice points perpendicular to its main diagonal.\\
The solution to this problem is known since long and traces back to Abraham de {\sc %%@
Moivre}\cite{ADM}:
\begin{equation}
\left|{\mathcal S}(A,N,M)\right|= 
\sum_{n=0}^{\lfloor\frac{M}{A}\rfloor}(-1)^n {N\choose n}{N-1+M-nA\choose N-1}, \label{G1}
\end{equation}
where $\lfloor x \rfloor$ denotes the largest integer $\le x$.
The proof is straight-forward using the generating function (see e.~g.~\cite{Fel68})
\begin{equation}
\left(\sum_{a=0}^{A-1}z^a \right)^N = \sum_{m=0}^{N(A-1)}\left|{\mathcal S}(A,N,m)\right|z^m. 
\end{equation}

%-----------------------------------------------------------
\section{Cycles of strings}

We will call the equivalence classes 
${\bf a}=[a]_\approx,\quad a\in{\mathcal S}(A,N)$
of strings which are connected by cyclic shifts ``cycles". The different sets of cycles will be %%@
denoted by
\begin{equation}
{\mathcal C}(A,N)\eqdef {\mathcal S}(A,N)/\approx,  \quad
{\mathcal C}(A,N,M)\eqdef {\mathcal S}(A,N,M)/\approx.
\end{equation}
This notation appears natural since cycles are the orbits of the cyclic group
\begin{equation}
G\eqdef \left\{ T^n :\, n=0,\ldots,N-1 \right\}\cong {\Bbb Z}_N
\end{equation}
operating on strings in the way defined above. Hence cycles can at most contain $N$ strings.
The number of strings contained in a cycle will be called its ``order".
``Proper cycles" are defined as those of maximal order $N$, ``epicycles" are cycles of order %%@
less than $N$. Special epicycles are those containing exactly one constant string $a=\langle %%@
i,i,\ldots,i\rangle, i\in\left\{ 0,\ldots, A-1 \right\}$. These will be of order one and are %%@
called ``monocycles". Obviously, there are exactly $A$ monocycles.\\
Generally, the orbit of a group $G$ generated by the operation on some element $a$ will be %%@
isomorphic to the quotient set $G/G_a$, where $G_a$ is defined as the subgroup of all %%@
transformations leaving $a$ fixed. In our case $G_a$ will be isomorphic to ${\Bbb Z}_k$ where $k$ %%@
is a divisor of $N$ and ${\bf a}$ will be of order $n=\frac{N}{k}$. The case $k=1$ corresponds %%@
to proper cycles, whereas the case $k=N$ yields monocycles.\\
To put it differently:
If a string $a\in{\mathcal S}(A,N)$ consists of $k$ copies of a substring $b\in{\mathcal S}(A,n)$, 
$k n = N$, it will generate an epicycle ${\bf a}=[a]_\approx$ containing at most $n$ strings. %%@
${\bf a}$ contains exactly $n$ strings iff $b$ itself generates a proper cycle 
${\bf b}\in {\mathcal C}(A,n)$. Conversely, any epicycle ${\bf a}$ of order $n$ consists of %%@
strings which are $k$ copies of substrings $b$ belonging to proper cycles ${\bf b}$. 
Moreover, if 
${\bf a}\in{\mathcal C}(A,N,M)$ is of order $n$ the corresponding proper cycle ${\bf b}$ will %%@
satisfy ${\bf b}\in {\mathcal C}(A,n,m)$ with $M=k m$. Thus we obtain the following
\begin{lemma}
\begin{enumerate}
\item The order $n$ of any cycle ${\bf a}\in{\mathcal C}(A,N,M)$ is a divisor of $N$.\\
\item Moreover, in this case $m\eqdef\frac{M n}{N}$ will be an integer.
\end{enumerate}
\end{lemma}
Hence the order of cycles will always belong to the following set:
\begin{Def}
${\mathcal D}(A,N,M)\eqdef \left\{ n\in{\Bbb N} :\, n|N \mbox{ and } N|M n \right\}$.
\end{Def}

In passing we note that if $N$ is a prime number, then there will be only proper cycles and
exactly $A$ monocycles, as mentioned above, hence $N$ will divide $A^N-A$, which is essentially
{\sc Fermat}'s theorem of 1640.\\

\begin{Def}
Let ${\mathcal N}(A,N,M,n)$ denote the number of cycles ${\bf a}\in {\mathcal C}(A,N,M)$ of order $n$
and ${\mathcal M}(A,N,M,n)$ the number of strings belonging to these cycles:
\begin{equation}
{\mathcal M}(A,N,M,n)\eqdef {\mathcal N}(A,N,M,n)n.
\end{equation}
\end{Def}

According to the preceding discussion the %%@
following holds:
\begin{lemma}
\begin{eqnarray}
|{\mathcal S}(A,N,M)|& =& \sum_{n\in{\mathcal D}(A,N,M)}{\mathcal M}(A,N,M,n),\label{G2}\\
 {\mathcal N}(A,N,M,n) & = & \left\{ \begin{array}{r@{\quad:\quad}l}
				{\mathcal N}(A,n,\frac{M n}{N},n) &\mbox{ if } n\in {\mathcal D}(A,N,M)\\
				0 & \mbox{ else } \end{array}\right..	
\end{eqnarray}
\end{lemma}
Together with (\ref{G1}) this yields a recursion relation for $ {\mathcal M}(A,N,M,n)$. It is, %%@
however, possible to obtain an explicit formula, which will be shown in the next section.

%-----------------------------------------------------------
\section{Explicit formula for $ {\mathcal M}(A,N,M,n)$}

Let us consider for example $N=12$. Then (\ref{G2}) yields the following equations, where redundant %%@
arguments will be suppressed:
\begin{eqnarray}
|{\mathcal S}(A,12,M)| & \eqdef & S_{12}\\
 & = & {\mathcal M}(A,12,M,12)+
	{\mathcal M}(A,6,M/2,6)+{\mathcal M}(A,4,M/3,4)
\nonumber\\
 & & +{\mathcal M}(A,3,M/4,3)+
	{\mathcal M}(A,2,M/6,2)+{\mathcal M}(A,1,M/12,1)\nonumber\\
 & \eqdef & M_{12}+M_6+M_4+M_3+M_2+M_1 \nonumber\\
 & = & M_{12}+(S_6-M_3-M_2-M_1)+(S_4-M_2-M_1)\nonumber\\
 & & +(S_3-M_1)+(S_2-M_1)+S_1\nonumber\\
 & = & M_{12}+(S_6-(S_3-S_1)-(S_2-S_1)-S_1)+\nonumber\\
 & & (S_4-(S_2-S1)-S_1)+(S_3-S_1)+(S_2-S_1)+S_1\nonumber\\
 & \Rightarrow & \nonumber\\
M_{12}&=& S_{12}-S_6-S_4+(1-1)S_3+(1+1-1)S_2+\nonumber\\
  & & (-1-1+1-1+1+1+1-1)S_1\nonumber
\ .
\end{eqnarray}
We see how each $S_n$ will enter in different ways into the expression for $M_{12}$
according to different ``divisor chains" $n|\ldots|N$. Here by a ``divisor chain" we understand
a finite sequence of numbers each of which is a divisor of the next one. In the example, there %%@
are ``odd" divisor chains $6|12$, $4|12$, $3|12$, $2|12$, $1|3|6|12$, $1|2|6|12$, $1|2|4|12$
(with an odd number of strokes $|$), %%@
and ``even" divisor chains $3|6|12$, $2|6|12$, $2|4|12$, $1|6|12$, $1|4|12$, $1|3|12$ and $1|2|12$.  %%@
Each even divisor chain $n|\ldots|12$ yields a term $+S_n$, each odd one a term $-S_n$ in the %%@
expression for $M_{12}$.\\

Generalizing this example, we conclude that
\begin{equation}
{\mathcal M}(A,N,M,N) = \sum_{n\in{\mathcal D}(A,N,M)} \Delta_{n,N}\cdot 
\left|{\mathcal S}(A,n,\frac{M n}{N})\right|
\end{equation}
where $\Delta_{n,N}$ is defined as the number of even divisor chains $n|\ldots|N$ minus the %%@
number of odd divisor chains $n|\ldots|N$.\\
Thus the problem is reduced to the task of finding an explicit formula for $\Delta_{n,N}$.
Obviously, $\Delta_{n,N}=\Delta_{1,N/n}\eqdef\Delta_{N/n}$ if $n|N$. Let $K=N/n$. Each divisor %%@
chain $k_0=1|k_1|k_2|\ldots|k_\mu=K$ corresponds in a $1:1$ manner to a factorization
of $K$ of the form $K=\frac{k_1}{1}\cdot\frac{k_2}{k_1}\cdots \frac{k_\mu}{k_{\mu-1}}\eqdef %%@
K_1\cdot K_2\cdot \ldots \cdot K_\mu$. Of course, permutations of different factors count as %%@
different factorizations since they give rise to different divisor chains. Even (resp.~odd)
divisor chains correspond to even (resp.~odd) $\mu$.
\begin{lemma}
If $K$ is a product of $\nu$ different primes, $K=p_1 p_2 \ldots p_\nu$, then %%@
$\Delta_K=(-1)^\nu$.
\end{lemma}
{\bf Proof}: We proceed by induction. If $K$ is prime, i.~e.~ $\nu=1$, there is only one odd %%@
(trivial) factorization $K=K_1$ and $\Delta_K=-1$.\\
Next we assume the formula to be valid for $K$ and are going to prove it for 
$K'=K\cdot p_{\nu+1}$, where $p_{\nu+1}$ is a prime different from $p_1,\ldots,p_\nu$.
Let $K=K_1\cdots K_\mu$ be an arbitrary factorization of $K$. There are two processes to %%@
obtain from this a factorization of $K'$: Multiplication of one of the $\mu$ factors by %%@
$p_{\nu+1}$, which does not alter the even/odd character of the factorization. The other %%@
process is insertion of $p_{\nu+1}$ into one of $\mu+1$ places. This yields a factorization of %%@
length $\mu+1$ and hence changes the  even/odd character. Obviously, every factorization of %%@
$K'$ will be obtained by exactly one of these two procedures. Denote by $O(K)$ the number of %%@
odd factorizations of $K$ and by $E(K)$ the number of even ones. Then the preceding argument %%@
shows that
\begin{equation}
E(K')=\mu E(K)+(\mu+1)O(K),
\end{equation}
and
\begin{equation}
O(K')=\mu O(K)+(\mu+1)E(K).
\end{equation}
Subtraction yields
$E(K')-O(K')=O(K)-E(K)$, hence
$\Delta_{K'}=-\Delta_K=(-1)^{\nu+1}$.
\hfill $\blacksquare$

\begin{lemma}
If in the prime factorization of $K$ at least one prime occurs twice or more, then %%@
$\Delta_K=0$.
\end{lemma}
{\bf Proof}: Let $K'=K\cdot p_{\nu+1}$ as in the preceding proof, but $p_{\nu+1}|K$. If %%@
$K'=K'_1\cdot K'_1\cdots K'_{\mu'}$ is any factorization of $K'$, it may be obtained from %%@
factorizations of $K$ by different processes of multiplication by or insertion of $p_{\nu+1}$. %%@
(For example, $12=3\cdot 2\cdot 2$ may be obtained from $6=3\cdot 2$ by insertion at two %%@
different places.) In order to make the process unique we make the convention to delete the %%@
leftmost occurrence of $p_{\nu+1}$ in $K'=K'_1\cdot K'_2\cdots K'_{\mu'}$ thereby arriving at a %%@
factorization of $K$. Vice versa, this means that we will only multiply or insert $p_{\nu+1}$ %%@
left from $K_\lambda$ (including $K_\lambda$ in the case of multiplication), if $K_\lambda$ is %%@
the first factor with $p_{\nu+1}|K_\lambda$. Hence $p_{\nu+1}$ can be multiplied with %%@
$\lambda$ factors and inserted at $\lambda$ places whence
\begin{equation}
E(K')=\lambda E(K)+\lambda O(K),
\end{equation}
and
\begin{equation}
O(K')=\lambda O(K)+\lambda E(K).
\end{equation}
Subtraction yields $E(K')=O(K')$ and thus $\Delta_{K'}=0$.
\hfill $\blacksquare$

In order to formulate our main result we define
\begin{equation}
q(\nu)\eqdef\left\{ \begin{array}{r@{\quad:\quad}l}
				 (-1)^m & \mbox{ if $\nu$ is a product of $m$ different primes,}\\
				0 & \mbox{ else } \end{array}\right. .
\end{equation}
Summarizing, we have proven the following
\begin{theorem}

\begin{equation}
{\mathcal M}(A,N,M,N)=\sum_{n\in{\mathcal D}(A,N,M)}q(\frac{N}{n})
\sum_{\nu=0}^{\lfloor \frac{M n}{N A}\rfloor}(-1)^\nu 
{n\choose \nu}{n-1+\frac{M n}{N}-\nu A \choose n-1},
\end{equation}
\begin{equation}
{\mathcal M}(A,N,M,n)=
\left\{ \begin{array}{r@{\quad:\quad}l}
 {\mathcal M}(A,n,\frac{M n}{N},n) & \mbox{ if } n\in{\mathcal D}(A,N,M) %%@
\\
0 & \mbox{ else} \end{array}\right.
\end{equation}
\end{theorem}

Let ${\mathcal M}(A,n)$ denote the number of strings belonging to cycles of order $n$, irrespective
of $M$. This number does not depend on the total length $N$ of the strings. By an analogous reasoning
as above we may conclude
\begin{theorem}

\begin{equation}
{\mathcal M}(A,n)=\sum_{k|n}q(\frac{n}{k})A^k.\label{G3}
\end{equation}
\end{theorem}

From this the number of cycles is obtained by division by $n$. 
Note that $n|{\mathcal M}(A,n)$, hence (\ref{G3}) generalizes {\sc Fermat}'s original result
to the case where $n$ need not be prime.\\
We close the article by giving 
a numerical example for $N=12$ and $A=5$ in table \ref{T-1-1}.

%-----------------------------------------------------------------------
\begin{table}[t]
\begin{center}
\begin{tabular}{|r|r|}\hline
Order $n$ & Number of cycles of order $n$ \\ \hline\hline
1 & 5\\
2 & 10\\
3 & 40\\
4 & 150\\
6 & 2580\\
12 & 20343700\\ \hline
\end{tabular}
\vspace*{5mm}
\end{center}
\caption{Number of cycles of order $n$ for $N=12$ and $A=5$.}\label{T-1-1} 
\end{table}
%----------------------------------------------------------------------- 

%%%%%%%%%%%%%%%%%%%%%%%%%%%%%%%%%%%%%%%%%%%%%%%%%%%%%%%%%%%%%%%%%%%%%%%%
\end{document}